\documentclass[12pt]{article}
\usepackage{times}
\usepackage{geometry}
\usepackage{xspace}
\usepackage[utf8]{inputenc}
\usepackage{enumitem,amssymb}
\usepackage{graphicx}
\usepackage{multicol}
\newlist{thematic}{itemize}{8}
\setlist[thematic]{label=$\square$}
\usepackage{pifont}
\def\apj{{\rm ApJ}}
\def\apjl{{\rm ApJL}}

\def\aap{{\rm A\&A}}

\def\nat{{\rm Nature}}

\newcommand\ltsima{$\; \buildrel <\over\sim \;$}
\newcommand\simlt{\lower.5ex\hbox{\ltsima}}
\newcommand\gtsima{$\; \buildrel >\over\sim \;$}
\newcommand\simgt{\lower.5ex\hbox{\gtsima}}

\newcommand\mearth {{M_\oplus}}

\newcommand{\Mstar}{\ensuremath{M_{\star}}\xspace}

\newcommand{\Mp}{\ensuremath{M_{P}}\xspace}

\newcommand{\sigjit}[1]{
        \ensuremath{
                \ifthenelse{\equal{#1}{}}{\sigma_\mathrm{jit}}{\sigma_\mathrm{jit,#1}}}
        \xspace
}

\newcommand{\zfree}[1]{\ensuremath{\ifthenelse{\isempty{#1}}{z_{\mathrm{free}}^{*}}{z_{\mathrm{free,#1}}}}\xspace}

\newcommand\etal{et~al.}

\begin{document}
\begin{flushleft}
\huge
Astro2020 Science White Paper \linebreak

Wide-Orbit Exoplanet Demographics \linebreak
\normalsize

\noindent \textbf{Thematic Areas:} \hspace*{60pt} $\boxtimes$ Planetary Systems \hspace*{10pt} $\boxtimes$ Star and Planet Formation \hspace*{20pt}\hfil\linebreak
$\square$ Formation and Evolution of Compact Objects \hspace*{31pt} $\square$ Cosmology and Fundamental Physics \linebreak
  $\square$  Stars and Stellar Evolution \hspace*{1pt} $\square$ Resolved Stellar Populations and their Environments \hspace*{40pt} \linebreak
  $\square$    Galaxy Evolution   \hspace*{45pt} $\square$             Multi-Messenger Astronomy and Astrophysics \hspace*{65pt}
 \linebreak
  
\textbf{Principal Author:}

Name:	David P.\ Bennett
 \linebreak						
Institution:  NASA Goddard Space Flight Center and University of Maryland
 \linebreak
Email: david.bennett@nasa.gov
 \linebreak
Phone:  (301) 286-5473
 \linebreak
 
\textbf{Co-authors:} 
Rachel Akeson$^{1}$,
Yann Alibert$^{2}$,
Jay Anderson$^{3}$,
Etienne Bachelet$^{4}$,
Jean-Phillipe Beaulieu$^{5,6}$,
Andrea Bellini$^{3}$,
Aparna Bhattacharya$^{7,8}$, 
Alan Boss$^{9}$,
Valerio Bozza$^{10}$,
Stephen Bryson$^{11}$,
Derek Buzasi$^{12}$,
Sebastiano Calchi Novati$^{1}$,
Jessie Christiansen$^{1}$,
Shawn D.~Domagal-goldman$^{7}$,
Michael Endl$^{13}$,
Benjamin J.\ Fulton$^{1}$,
Calen B.~Henderson$^{1}$,
B.~Scott Gaudi$^{14}$,
Samson A.~Johnson$^{14}$,
Naoki Koshimoto$^{5,6,15}$,
Michael Meyer$^{16}$,
Gijs D.~Mulders$^{17}$,
Susan Mullally$^{3}$,
Ruth Murray-Clay$^{18}$,
David Nataf$^{19}$,
Eric Nielsen$^{20}$,
Henry Ngo$^{21}$,
Ilaria Pascucci$^{22}$,
Matthew Penny$^{14}$,
Peter Plavchan$^{23}$,
Radek Poleski$^{14}$,
Cl\'ement Ranc$^{7}$,
Sean N. Raymond$^{24}$,
Leslie Rogers$^{18}$,
Johannes Sahlmann$^{3}$,
Kailash C.~Sahu$^{3}$,
Joshua Schlieder$^{7}$,
Yossi Shvartzvald$^{25}$,
Alessandro Sozzetti$^{26}$,
Rachel Street$^{4}$, 
Takahiro Sumi$^{27}$, 
Daisuke Suzuki$^{28}$,
Neil Zimmerman$^{7}$
\vspace{0.3cm}
\begin{multicols}{2}
\noindent $^{1}$IPAC/Caltech \\
$^{2}$University of Bern, Switzerland\\
$^{3}$Space Telescope Science Institute \\
$^{4}$Las Cumbres Observatory \\
$^{5}$University of Tasmania, Australia \\
$^{6}$Institut d'Astrophysique de Paris, France \\
$^{7}$NASA Goddard Space Flight Center \\
$^{8}$University of Maryland \\
$^{9}$Carnegie Institution\\
$^{10}$University of Salerno, Italy \\
$^{11}$NASA Ames Research Center \\
$^{12}$Florida Gulf Coast University \\
$^{13}$University of Texas\\
$^{14}$Ohio State University \\
$^{15}$University of Tokyo \\
$^{16}$University of Michigan \\
$^{17}$University of Chicago \\
$^{18}$University of California, Santa Cruz \\
$^{19}$Johns Hopkins University \\
$^{20}$Stanford University \\
$^{21}$National Research Council, Canada\\
$^{22}$University of Arizona \\
$^{23}$George Mason University\\
$^{24}$Laboratoire d'Astrophysique de Bordeaux\\
$^{25}$Jet Propulsion Laboratory \\
$^{26}$INAF - Osservatorio Astrofisico di Torino\\
$^{27}$Osaka University, Japan \\
$^{28}$ISAS, JAXA, Japan \\
\end{multicols}

\textbf{Abstract:}
The Kepler, K2 and TESS transit surveys are revolutionizing our understanding of planets orbiting close
to their host stars and our understanding of exoplanet systems in general, but there remains a
gap in our understanding of wide-orbit planets. This gap in our understanding must be filled
if we are to understand planet formation and how it affects exoplanet habitability.
We summarize current and planned exoplanet detection programs using a variety of 
methods: microlensing (including WFIRST), radial velocities, Gaia astrometry, and direct imaging.
Finally, we discuss the prospects for joint analyses using results from multiple methods and
obstacles that could hinder such analyses.

We endorse the findings and recommendations published in the 2018 National 
Academy report on Exoplanet Science Strategy. This white paper extends and complements 
the material presented therein.

\end{flushleft}
\pagebreak
\section{The Need for Exoplanet Demographics}
\vspace{-0.3cm}
When, where, and how frequently do habitable planets form in our galaxy and throughout the 
observable Universe?  This question motivates a significant fraction of exoplanet, and indeed 
astronomical research today, in order to help us address whether or not we are alone in the Universe.  
In the past quarter century, we have made dramatic strides in our understanding of planetary systems,
and we have made great progress in the development of instrumentation with the capability of detecting
signs of life in planetary systems beyond our own. Nevertheless, we should keep in mind that our 
ignorance still greatly exceeds our knowledge of planet formation and especially planetary habitability. 
The conventional definition of the habitable zone is largely based on observations in our own solar system, and its 
position might vary for atmospheres very different from our own (e.g.\ Seager 2013). Furthermore, 
wide-orbit planets can influence the habitability of Earth-like planets in Earth-like orbits. 
The early formation of Jupiter's core could be one reason why the Earth does not have orders of magnitude 
more water (Morbidelli \etal\ 2016) in the context of pebble accretion scenarios. 
The small amount of water on the Earth could also result from the accretion of planetesimals formed interior 
to the snow line (e.g.\ Alibert et al. 2013).
Also, the growth and migration 
of wide-orbit planets invariably scatters water-rich planetesimals across their planetary systems, which 
provide a sprinkling of water to rocky planets (Walsh et al 2011; Alibert et al. 2013; Raymond \& Izidoro 2017).

Planet formation involves a multitude of complex processes that we do not have the ability to
simulate in detail. As a result, observations have regularly discovered planetary systems with
unexpected properties, such a hot Jupiters (Mayor \& Queloz 1995) systems with numerous,
low-density planets in very short period orbits (Lissauer \etal\ 2011). Thus, our understanding
of planetary systems is driven by observations, with theory as a critical tool for interpreting 
the observations. We have made great progress in understanding planets in close orbits with
radial velocities (e.g.\ Mayor \etal\ 2011) and especially the Kepler transit survey (Thompson \etal\ 2018),
but our understanding of wide-orbit planets is lacking. In this white paper, we describe how 
a multitude of methods can be used to advance our understanding of wide-orbit 
exoplanet demographics.

\begin{figure}[h]
 \begin{minipage}[t]{0.70\textwidth}
  \includegraphics[width=\textwidth]{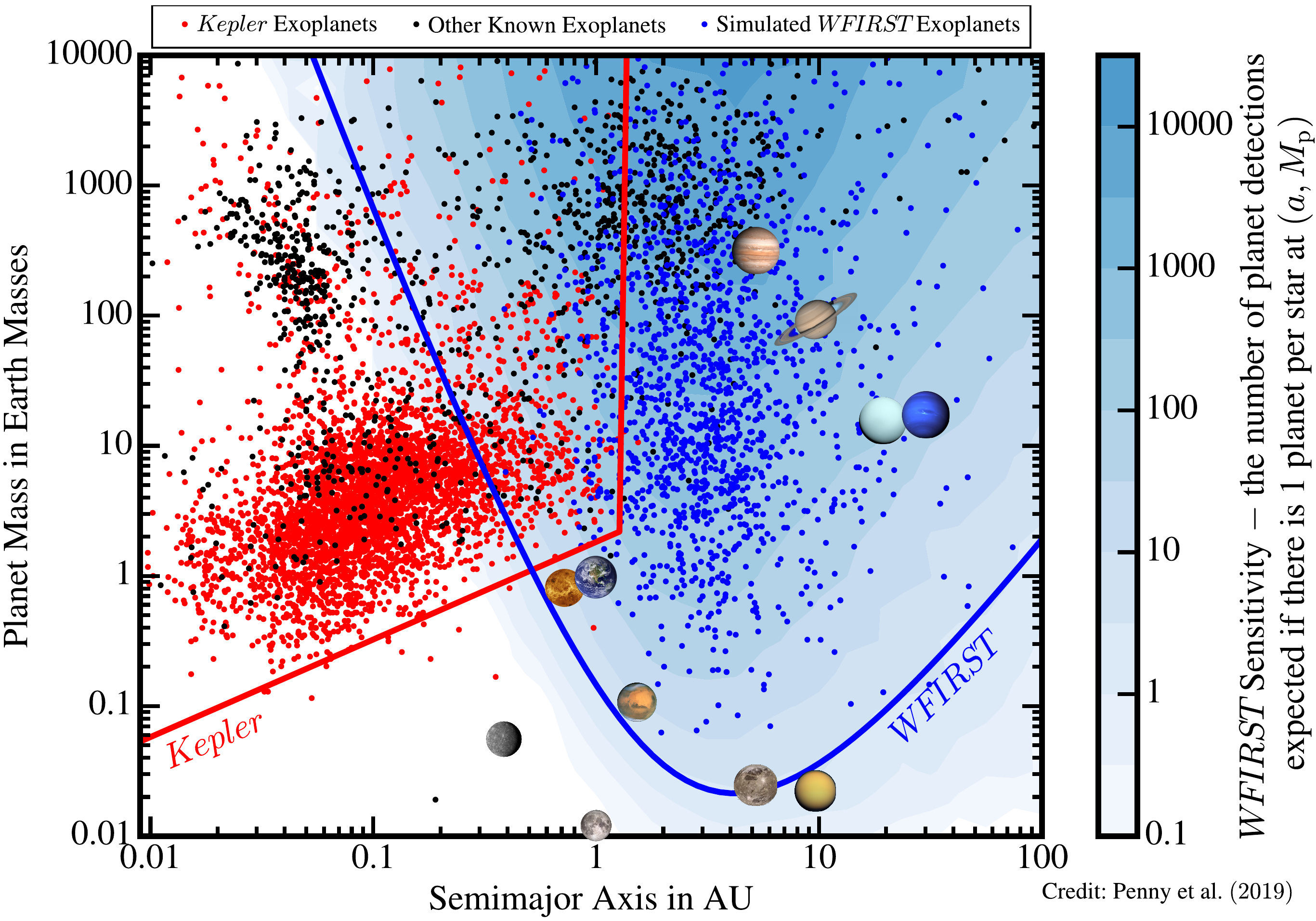}
 \end{minipage}\hfill
 \begin{minipage}[b]{0.26\textwidth}
\caption{Comparison of planet mass and semimajor axis for planets candidates found by Kepler (red), 
planet discoveries by other current methods (black) and simulated planets from WFIRST (blue). 
The pictures show the locations of solar system
planets and moons (as if they are planets).}
 \end{minipage}
\end{figure}

\vspace{-0.3cm}
\section{Microlensing with WFIRST and Ground-based Programs}
\vspace{-0.3cm}

The greatest planned investment in exoplanet demographics is the microlensing survey of
the WFIRST mission. Fig.~1 compares the expected WFIRST exoplanet discoveries with 
Kepler's planet candidates and planets found by other methods (Penny \etal\ 2019).
WFIRST's space-based microlensing survey was selected by the Astro2010 decadal 
survey because of the high sensitivity of space-based microlensing to low-mass planets
in wide orbits, ranging from the habitable zone of FGK stars to infinity, i.e.\ unbound planets
(Mr{\'o}z \etal\ 2019).
As Fig.~1 indicates, WFIRST is sensitive to analogs of all the planets in our Solar System,
except for Mercury, and has sensitivity extending down to planets below the mass of Mars 
($\simlt 0.1\mearth$). This is more than 2 orders of magnitude lower than the sensitivity 
of other methods.

While microlensing light curves themselves usually yield only the star-planet mass ratio, $q$, WFIRST's 
space-based imaging will yield host star and planet masses for the majority of planets
discovered using a combination of microlensing light curve constraints and direct observations
of the host star (Bennett \etal\ 2007; Bhattacharya \etal\ 2018). This also yields the
lens systems' distance. Microlensing is also sensitive to
planets in binary systems (Gould \etal\ 2013; Bennett \etal\ 2016), and WFIRST will be sensitive
to exomoons in systems similar to the Earth-moon system (Bennett \& Rhie 2002).

\vspace{-0.2cm}
\begin{figure}[h]
\includegraphics[width=3.0in]{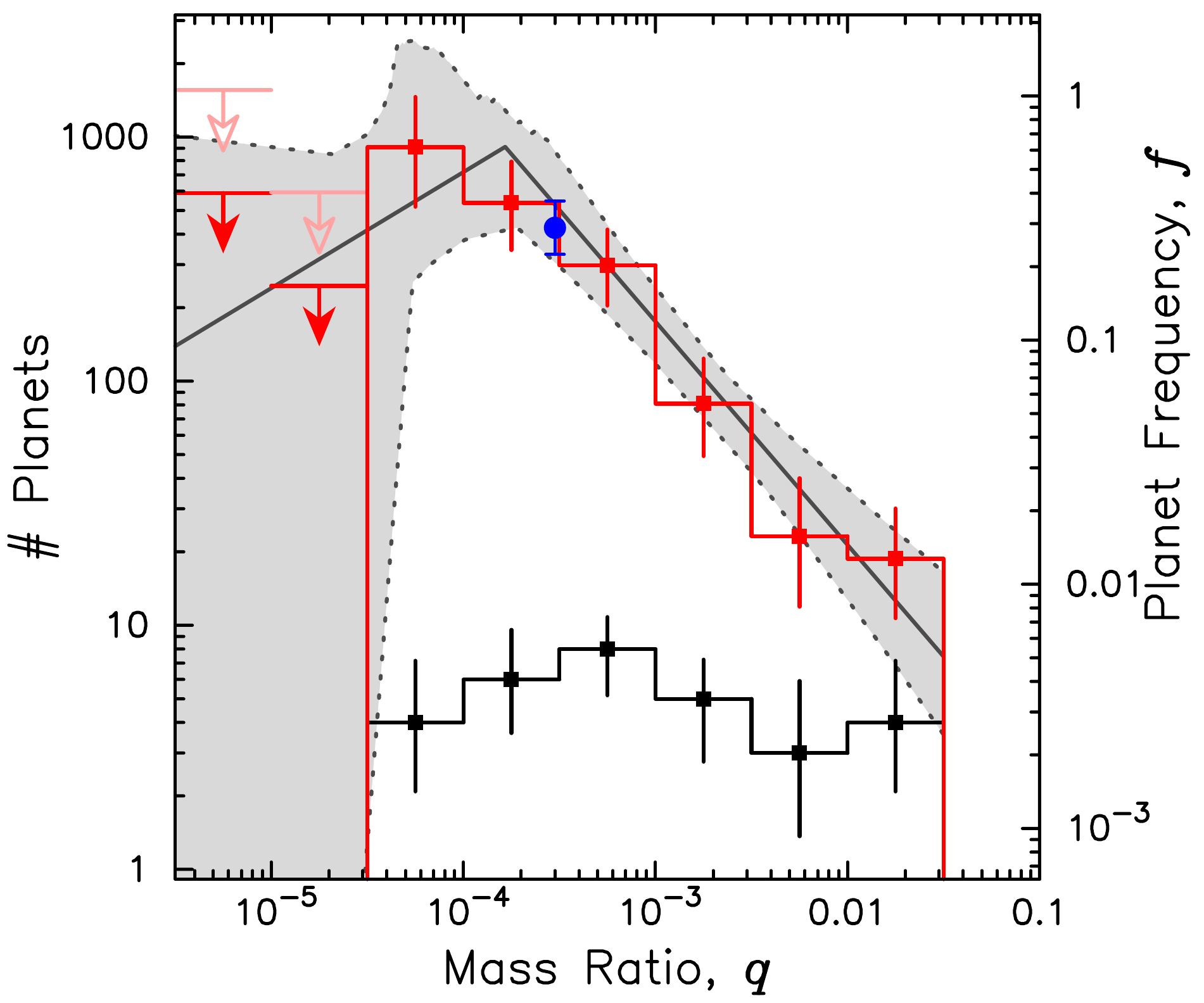}\ \ \ \ \ 
\includegraphics[width=3.2in]{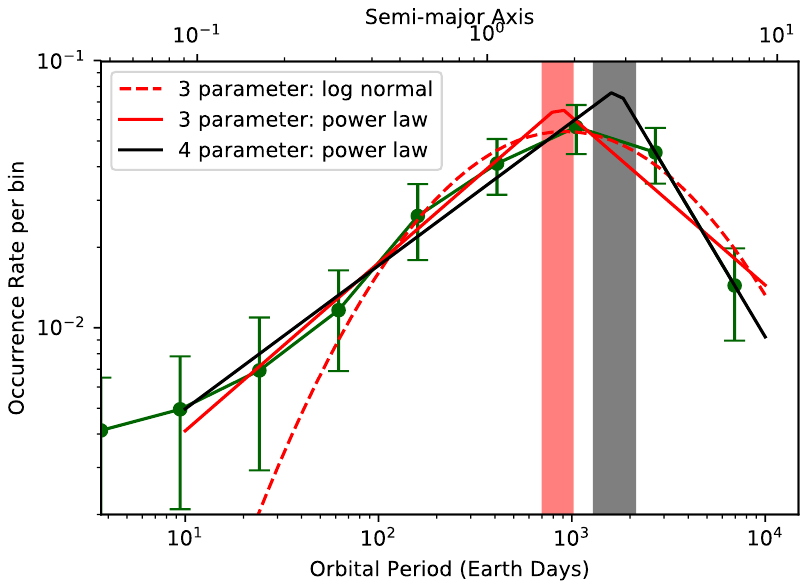}
\vspace{-0.2cm}
\caption{The left panel shows the raw microlens planet mass ratio distribution (Suzuki \etal\ 2016) 
in black, with the detection efficiency corrected distribution in red. The grey shading
indicates the range of 1$\sigma$ broken power law models. The peak at $q\approx 10^{-4}$
has been confirmed by Jung \etal\ (2019). The right panel shows the gas giant planet occurrence rate distribution
for a large radial velocity sample (Fernandes \etal\ 2018) indicating a peak at periods 1000-2000 days.}
\end{figure}
\vspace{-0.2cm}

Prior to WFIRST, ground-based microlensing surveys will provide the highest sensitivity to low mass planets
in wide orbits around GKM stars. 
Recent results from the MOA group (see Fig.~2) show a smooth distribution of planets from
mass ratio $q = 0.03$ down to $q \approx 6\times 10^{-5}$ in seeming contradiction to a gap at $1$-$4\times 10^{-4}$
(or $20$-$80\mearth$) predicted by the runaway accretion scenario of the core accretion model (Suzuki \etal\ 2018).
The 3-telescope network of the KMTNet survey (Jung \etal\ 2019) is now detecting planets at a 
much higher rate than MOA, and
should improve these statistics significantly prior to WFIRST. These ground-based microlensing
demographic results can help to enable the
planning of other near term programs, such as the search for sub-Saturn and Neptune mass exoplanets 
orbiting young M-dwarfs with JWST (Schlieder 2019).

\vspace{-0.3cm}
\section{Long-period Exoplanet Demographics with Radial Velocities}
\vspace{-0.3cm}

The radial velocity (RV) method has sensitivity to planets at a very large range of orbital periods ranging
from $\sim 1\,$day to $>20\,$years, which corresponds to a factor of $\sim 400$ in semi-major axes.
In recent years, much of the RV effort has been focused on observations of transiting planets found
by Kepler, K2, and now, TESS (e.g.\ Marcy \etal\ 2014; Petigura \etal\ 2017; Gandolfi \etal 2018; Palle \etal\ 2019). 
Much of the effort in the near future will be focused on improving
the precision of RV measurements for planets in the vicinity of the habitable zone (Fischer \etal\ 2016).

The amplitude of RV signal (K) caused by an orbiting planet is 
$K \propto \Mp \,\Mstar^{-1/2} a^{-1/2}$,
where \Mp is the mass of the planet, \Mstar is the mass of the host star, and $a$ is semi-major axis,
so the sensitivity degrades fairly gradually with semi-major axis. However, there are additional challenges
for detecting long-period planets with the RV method. The planetary orbital periods of these planets can
be similar to stellar magnetic cycles, so care must be taken to separate the RV and magnetic cycle effects.
This requires some more observing time, but it can be done with RV data (Endl \etal\ 2016).
Also, these periods are longer than healthy PhD timescales, and it
can be a challenge to maintain records of observing decisions that might be needed for statistical
analyses for planets with periods measured in decades. Statistical analyses of RV data can be 
compromised if observing plans are changed due to the detection of candidate planetary signals
when a fraction of the data has been collected. Therefore, it would be useful to establish
a repository of long duration data sets, including the details of any decisions to change observing 
plans to enable to RV studies of planets with multi-decade orbital periods.

Fortunately, several long duration surveys have been completed 
or are still ongoing (e.g., Mayor \etal\ 2011, Fischer \etal\ 2016, Howard \etal\ 2016)
but with a plethora of new RV instruments coming online (Fischer \etal\ 2016)
we need to conduct simultaneous surveys on both the old and new instruments in order to maximize 
the value of the legacy RV datasets and to maximize the observing baselines of new surveys with the 
latest generation of extremely-precise RV instruments. 

The Mayor \etal\ (2011) sample of FGK stars has been analyzed by Fernandes \etal\ (2018) who found a peak in the orbital
period distribution at periods of 1000-2000 days for planets of $>50\mearth$, 
as shown in the right panel of Fig.~2. This is consistent with the results of direct detection surveys, but
it has not yet been compared in detail to the Suzuki \etal\ (2016) microlensing results.

\vspace{-0.4cm}
\section{Astrometry from Gaia}
\vspace{-0.3cm}

ESA's ongoing Gaia mission is expected to release a large catalog of planet discoveries in the final
data release from its prime mission in 2022 or 2023 based on astrometric data from more than a million 
FGKM stars. The detection of $\sim 20,000$ planets of Jupiter mass or larger at orbital
separations of $0.5$ to $5\,$AU is expected (Perryman \etal\ 2014), and the 2-dimensional astrometric
orbits will provide planetary mass measurements for the detected planets. The Gaia mission has
already been extended to a 6.5 year duration, and based on 
Gaia's success to date, it seems likely that its mission will be extended to a full 10 years . 
This might triple the number of detected planets
and detect as many as 3000 systems with two giant planets (Casertano \etal\ 2008). 
For $\sim$10\% of these systems, sufficiently accurate orbital solutions would allow for 
precise mutual inclination angle determinations. Gaia's huge sample
will allow a detailed investigation of the diversity of giant exoplanets in wide orbits.

The Gaia results will complement the RV studies quite well. Gaia will provide the inclinations 
and masses for previously discovered RV planets, while the RV data will provide information on 
planets orbiting close to their host stars that are beyond Gaia's sensitivity.
Gaia's fantastic statistics on giant planets orbiting stars in the Galactic neighborhood can be 
compared to the statistics of giant planets with measured distances from the WFIRST microlensing
survey to determine how the population of wide-orbit giant planets depends on the Galactic environment.

\vspace{-0.4cm}
\section{Direct Imaging}
\vspace{-0.3cm}

Direct imaging is complementary to the indirect techniques discussed in this paper. Infrared imaging
provides estimates of planet temperature and luminosity (and thus radius), as well as atmospheric 
composition, while optical imaging can provide information on rotation rate, dynamics, and climate.
Such information provides fundamental constraints on possible habitability as well.  
Planets can be directly detected in reflected light, or in thermal emission, either from the residual heat 
of formation or in thermal equilibrium with the host star.  It is easiest to image self-luminous planets farther 
from their host stars as observations are currently limited by the relative contrast achievable (typically $10^{-6}$ at 1").  

Current surveys are capable of detecting young self-luminous gas giant planets at large 
orbital radii (e.g. $> 1 M_{\rm Jup}$ at $> 30 $AU, where $ M_{\rm Jup}$ is Jupiter's mass). 
Nearby young stars are rare, so host samples 
often use stars in nearby moving groups with ages of 30-300 Myr (providing better mass sensitivity) 
at distances of 20-200 pc, with the physical resolution depending on the distance.  
These surveys are conducted with 6-12 meter telescopes employing high contrast imaging systems 
utilizing high actuator density adaptive optics systems (e.g. GPI, SPHERE, MagAO,  SCExAO).  
Recent work suggests that gas giant planets $1-10 M_{\rm Jup}$ are rare beyond 30 AU 
(Nielsen et al. submitted; Vigan et al. in preparation).  Combined with data from other techniques, this 
implies a planet surface density distribution that rises to a peak between 1-10 AU, and 
then falls for both M dwarf and FGK star populations (Clanton \& Gaudi 2016; Meyer et al. 2018; 
Fernandes et al. 2019).  Sensitive gas giant planet searches also suggest a local minimum in the 
companion mass function between $10-40M_{\rm Jup}$ (Reggiani et al. 2016). To date, the measured 
frequency of gas giant planets through direct imaging appears to depend on stellar mass in that higher mass stars 
have a higher giant planet occurrence rate (e.g. Bowler 2016), which is the opposite of the case
for lower mass planets in short period orbits 
(Howard \etal\ 2012; Petigura \etal\ 2013; Mulders \etal\ 2015; Dressing \etal\ 2015).

Future work with the next generation extremely large telescopes (ELTs) will enable surveys to probe to 
smaller inner working angles ($<  3\,$AU) with greater sensitivity ($< 0.5 M_{\rm Jupr}$) around nearby young stars).  
The James Webb Space Telescope (JWST) will have extraordinary sensitivity from 1-28 microns compared 
to the ground when operating in the background limit ($> 1$"), but  it will likely be inferior to ground-based 
AO in the contrast limit; (cf. Danielski et al. 2018; Delacroix et al. 2013). In the background limit 
($> 20\,$AU for typical targets), JWST should be able to detect planets as small as Uranus and Neptune 
if they are common. Both ground-based AO on current and future telescopes as well as HST and JWST 
characterization of wide-orbit companions will enable reliable assessment of planet atmospheric 
composition (e.g.\ volatiles such as C/O ratio) compared to the host star.  Such studies provide fundamental 
tests of planet formation theory, both origin location and migration history.  

A major breakthrough is expected with ELTs as they should be able to image small planets ($1$-$4 R_\oplus$) 
around the very nearest stars in both reflected light and thermal emission.  Results from these surveys, 
when combined with those from other techniques, will enable us to assess: i) discontinuities in 
planet mass functions and orbital surface density distributions and how they may depend 
on each other; ii) how diverse planet compositions depend on mass, radius, and orbital location; 
and iii) how all of the above depend on host star properties such as composition, multiplicity, 
and formation environment (e.g. field versus open cluster).  Addressing these questions will 
enable us to put predictive planet formation models to the test, allowing us to take the next 
steps to assessing the frequency of planetary environments that may give rise to the 
biochemical origins of life. 

\vspace{-0.2cm}
\section{Joint Analysis of Exoplanet Samples from Different Methods}
\vspace{-0.2cm}

This white paper grew out of discussions in the Science Interest Group (SIG) \#2 of the
Exoplanet Exploration Program Analysis Group (ExoPAG), which focuses on ``Exoplanetary System 
Demographics\rlap," led by Jessie Christiansen and Michael Meyer.
This group contains practitioners of all the exoplanet detection methods 
outlined in this paper, but in our discussions it became apparent many, if not most, SIG \#2 members
were unaware of the latest developments using other methods. However, it is also clear that our
knowledge of exoplanet demographics can be advanced by combining the results from multiple 
methods (Clanton \& Gaudi 2014a, 2014b; Pascucci \etal\ 2018), 
but this can become difficult if the practitioners of each method are unfamiliar with the details
of the other methods. For example, the microlensing and radial velocity studies shown in Fig.~2 are
not easily combined. The microlensing analysis was done for a sample of GKM stars using planet-star mass ratios 
because the host star masses are not always known, but the RV analysis was done using planet masses.
While we know the masses of the RV planet host stars, we cannot convert the analysis to use mass ratios,
because the detection efficiencies for the individual stars in the survey are not available. Fortunately,
there is an ongoing program (Bhattacharya \etal\ 2018) to determine the masses of the host stars in the 
Suzuki \etal\ (2016) sample, so it will soon be possible to do a joint analysis using planet and host
masses with the Mayor \etal\ (2011) sample. 

We expect that this sort of problem will be very common when comparing demographic results
from different studies. In fact, it can occur with studies using the same method. If the samples
overlap, then some adjustment needs to be made to avoid over-counting the stars in the 
sample that overlap. This generally requires detection efficiencies for individual stars.
Therefore, we recommend that a national exoplanet demographics database be established to
collect detailed demographic data, like individual star exoplanet detection efficiencies, so that
the results of published demographic studies can easily be incorporated into future studies.

\pagebreak
\noindent\textbf{References}\\
Alibert, Y., Carron, F., Fortier, A., et al.\ 2013, \aap, 558, A109\\
Bennett, D.~P., Anderson, J., \& Gaudi, B.~S.\ 2007, ApJ, 660, 781 \\
Bennett, D.~P., \& Rhie, S.~H.\ 2002, ApJ, 574, 985 \\
Bennett, D.~P., Rhie, S.~H., Udalski, A., et al.\ 2016, AJ, 152, 125 \\
Bhattacharya, A., Beaulieu, J.-P., Bennett, D.~P., et al.\ 2018, AJ, 156, 289 \\
Bowler, B.~P.\ 2016, PASP, 128, 102001 \\
Casertano, S., Lattanzi, M.~G., Sozzetti, A., et al.\ 2008, A\&A, 482, 699 \\
Clanton, C., \& Gaudi, B.~S.\ 2014a, \apj, 791, 90 \\
Clanton, C., \& Gaudi, B.~S.\ 2014b, \apj, 791, 91 \\
Clanton, C., \& Gaudi, B.~S.\ 2016, ApJ, 819, 125 \\
Dressing, C.~D., \& Charbonneau, D.\ 2015, \apj, 807, 45 \\
Endl, M., Brugamyer, E.~J., Cochran, W.~D., et al.\ 2016, \apj, 818, 34\\
Fernandes, R.~B., Mulders, G.~D., Pascucci, I., Mordasini, C., \& Emsenhuber, A.\ 2018, arXiv:1812.05569 \\
Fischer, D.~A., Anglada-Escude, G., Arriagada, P., et al. 2016, PASP, 128, 066001 \\
Gandolfi, D., Barrag{\'a}n, O., Livingston, J.~H., et al.\ 2018, \aap, 619, L10 \\
Gould, A., Udalski, A., Shin, I.-G., et al.\ 2014, Science, 345, 46 \\
Howard, A.~W., Marcy, G.~W., Bryson, S.~T., et al.\ 2012, ApJS, 201, 15 \\
Jung, Y.~K., Gould, A., Zang, W., et al.\ 2019, AJ, 157, 72  \\
Lissauer, J.~J., Fabrycky, D.~C., Ford, E.~B., et al.\ 2011, Nature, 470, 53 \\
Morbidelli, A., Bitsch, B., Crida, A., et al.\ 2016, Icarus, 267, 368\\ 
Marcy, G.~W., Isaacson, H., Howard, A.~W., et al.\ 2014, ApJS, 210, 20 \\
Mayor, M., Marmier, M., Lovis, C., \etal\ 2011, arXiv:1109.2497 \\
Mayor, M., \& Queloz, D. 1995, Nature, 378, 355\\
Meyer, M.~R., Amara, A., Reggiani, M., \& Quanz, S.~P.\ 2018, \aap, 612, L3 \\
Mr{\'o}z, P., Udalski, A., et al.\ 2019, \aap, 622, A201\\
Mulders, G.~D., Pascucci, I., \& Apai, D.\ 2015, \apj, 814, 130 \\
Palle, E., Nowak, G., Luque, R., et al.\ 2019, \aap, 623, A41 \\
Pascucci, I., Mulders, G.~D., Gould, A., \& Fernandes, R.\ 2018, \apjl, 856, L28 \\
Penny, M.~T., Gaudi, B.~S., Kerins, E., et al.\ 2019, ApJS, 241, 3 \\
Perryman, M., Hartman, J., Bakos, G.~{\'A}., \& Lindegren, L.\ 2014, ApJ, 797, 14 \\
Petigura, E.~A., Marcy, G.~W., \& Howard, A.~W.\ 2013, \apj, 770, 69 \\
Petigura, E.~A., Sinukoff, E., Lopez, E.~D., et al.\ 2017, AJ, 153, 142 \\
Raymond, S.~N., \& Izidoro, A.\ 2017, Icarus, 297, 134 \\
Reggiani, M., Meyer, M.~R., Chauvin, G., et al.\ 2016, \aap, 586, A147 \\
Schlieder, J., 2019, JWST GTO Proposal 1184 \\
Seager, S., 2013, Science, 340, 577 \\
Sozzetti, A., \& de Bruijne, J.\ 2018, Handbook of Exoplanets, 81 \\
Suzuki, D., Bennett, D.~P., Ida, S., et al.\ 2018, ApJL, 869, L34 \\
Suzuki, D., Bennett, D.~P., Sumi, T., \etal\ 2016, ApJ, 833, 135 (S16) \\
Thompson, S.~E., Coughlin, J.~L., Hoffman, K., et al.\ 2018, ApJS, 235, 38 \\
Walsh, K.J., Morbidelli, A., Raymond, S.N., O'Brien, D.P., Mandell, A.M.\ 2011, \nat, 475, 206 \\
Yee, J.~C., \etal\ 2019, Astro2020 survey white paper \\

\end{document}